\documentclass[12pt]{article}

\usepackage{epsfig,amsmath,amssymb,graphics,graphicx} 

\newcommand{\be}{\begin{equation}}
\newcommand{\ee}{\end{equation}}
\newcommand{\bi}[1]{\vspace{-3mm} \bibitem{#1}}

\def\bfkappa{\mathop{\mbox{\boldmath $\kappa$}}}

\begin{document}

\begin{center}
{\it Journal of Mathematical Physics 47 (2006) 092901}
\end{center}

\begin{center}
{\Large \bf Map of Discrete System into Continuous} 
\vskip 5 mm

{\large \bf Vasily E. Tarasov} \\

\vskip 3mm

{\it Skobeltsyn Institute of Nuclear Physics, \\
Moscow State University, Moscow 119991, Russia } \\
{E-mail: tarasov@theory.sinp.msu.ru}
\end{center}

\vskip 11 mm

\begin{abstract}
Continuous limits of discrete systems
with long-range interactions are considered. 
The map of discrete models into continuous medium models is defined. 
A wide class of long-range interactions that
give the fractional equations in the continuous limit is discussed.
The one-dimensional systems of coupled oscillators for 
this type of long-range interactions are considered.
The discrete equations of motion are mapped into 
the continuum equation
with the Riesz fractional derivative. 
\end{abstract}


\vskip 11 mm

\section{Introduction}

Equations which involve derivatives or integrals of noninteger order 
\cite{SKM,OS,MR,Podlubny,KST}
have found many applications in recent studies in mechanics and physics 
\cite{Zaslavsky2,Zaslavsky1,Mainardi,Hil,Tar1,Tar2}. 
Usually the fractional equations for dynamics or kinetics 
appear as some phenomenological models. 
Recently, the method to obtain fractional analogues of equations 
of motion was considered for sets of coupled particles 
with a long-range interaction \cite{LZ,TZ3,KZT}. 
Examples of systems with interacting oscillators, spins, or waves 
are used for numerous applications 
in physics, chemistry, biology 
\cite{Dyson,J,NakTak,S,CMP,Kur,Ruf,BK,PV,Br4,GF,LLI}. 
Transfer from the equations of motion for discrete systems
to the continuous medium equation 
with fractional derivatives is an approximate procedure. 
Different applications of the procedure have already been used 
to derive fractional sine-Gordon and fractional wave 
Hilbert equation \cite{LZ,KZT}, to study synchronization of 
coupled oscillators \cite{TZ3}, and for fractional 
Ginzburg-Landau equation \cite{TZ3}.

Long-range interaction has been the subject 
of great interest for a long time. 
Thermodynamics of the model of classical spins with long-range
interactions has been studied in Refs. \cite{Dyson,J,CMP,NakTak}.
An infinite one-dimensional Ising model with long-range interactions 
was considered by Dyson \cite{Dyson}.
The $d$-dimensional classical Heisenberg model with long-range 
interaction is  described in Refs. \cite{J,CMP}, and
their quantum generalization 
can be found in Ref. \cite{NakTak}.
The long-range interactions have been 
widely studied in discrete systems on lattices 
as well as in their continuous analogues.  
Solitons in a one-dimensional lattice with the 
long-range Lennard-Jones-type interaction were considered in Ref. \cite{Ish}. 
Kinks in the Frenkel-Kontorova model with long-range 
interparticle interactions were studied in Ref. \cite{BKZ}. 
The properties of time periodic spatially localized solutions (breathers) 
on discrete chains in the presence of algebraically decaying interactions 
were considered in Refs. \cite{Br4,GF}. 
Energy and decay properties of discrete breathers in systems with long-range 
interactions have also been studied in the framework of the Klein-Gordon 
\cite{BK}, and discrete nonlinear Schrodinger equations \cite{Br6}. 
A remarkable property of the dynamics described by the equation with 
fractional space derivatives is that the 
solutions have power-like tails. 
Similar features were observed 
in the lattice models with power-like long-range 
interactions \cite{PV,Br4,GF,AEL,AK,APV,KZT}. 
As it was shown in Ref. \cite{TZ3,KZT}, 
analysis of the equations with fractional derivatives can provide 
results for the space asymptotics of their solutions. 

The goal of this paper is to study a connection between  
the dynamics of system of particles with long-range interactions 
and the fractional continuous medium equations 
by using the transform operation. 
Here, we consider the one-dimensional lattice 
of coupled nonlinear oscillators. 
We make the transform to the continuous limit 
and derive the fractional equation which describes 
the dynamics of the oscillatory medium.
We show how the continuous limit for the systems 
of oscillators with long-range interaction
can be described by the corresponding 
fractional equation.

In Sec. 2, the equations of motion for the system 
of oscillators with long-range interaction are considered.
In Sec. 3, the transform operation that maps the discrete
equations into continuous medium equation is defined.
In Sec. 4, the Fourier series transform of the equations of 
a system with long-range interaction is realized.
In Sec. 5, we consider a wide class of long-range interactions 
that can give the fractional equations in the continuous limit.
In Sec. 6, the simple example of nearest-neighbor interaction
is considered to demonstrate the application of the transform
operation to the well-known case.
In Sec. 7, the power-law long-range interactions with 
positive integer powers are considered.
In Sec. 8, the power-law long-range interactions with 
noninteger powers and the correspondent continuous medium equations
are discussed.
In Sec. 9, the nonlinear long-range interactions for
the discrete systems are used to derive the Burgers, 
Korteweg-de Vries and Boussinesq equations 
and their fractional generalizations in the continuous limit. 
In Sec. 10, the fractional equations are obtained from
the dispersion law for three-dimensional discrete system.
The conclusion is given in Sec. 11.


\section{Equations of motion for interacting oscillators}

Consider a one-dimensional system of 
interacting oscillators that are described by 
the equations of motion,
\be \label{Main_Eq}
\frac{\partial^2 u_n}{\partial t^2} = g  
\hat {\cal I}_{n} (u)  + F (u_n) ,
\ee
where $u_n$ are displacements from the equilibrium. 
The terms $F(u_n)$ characterize an interaction of the oscillators   
with the external on-site force. 
The term $\hat {\cal I}_{n}(u)$ is defined by
\be \label{Z3}
\hat {\cal I}_{n} (u) \equiv 
\sum_{\substack{m=-\infty \\ m \ne n}}^{+\infty} \; 
J(n,m) \; E(u_n,u_m) ,
\ee
and it takes into account the interaction 
of the oscillators in the system. \\

{\bf Examples}.  \\
1) If $J(n,m)=\delta_{n+1, m}-\delta_{n,m}$, and $E(u_n,u_m)=u_m$,
then $\hat {\cal I}_{n}(u) =u_{n+1}-u_n =\Delta u_n$. \\
2) For $J(n,m)=\delta_{n+1, m}-2\delta_{n,m}+\delta_{n-1,m}$,
and $E(u_n,u_m)=u_m$, we get
\[ \hat {\cal I}_{n} (u) =u_{n+1}-2u_n+u_{n-1}=\Delta^2 u_n . \]
3) We can consider the long-range interaction that is given by 
$J(n)=|n|^{-(1+\alpha)}$, where $\alpha$ is a positive real number.
In this  case, we have nonlocal coupling 
given by the power-law function. 
Constant $\alpha$ is a physical relevant parameter.
Some integer values of $\alpha$ correspond to 
the well-known physical situations: 
Coulomb potential corresponds to $\alpha=0$, dipole-dipole interaction 
corresponds to $\alpha=2$, and the limit $\alpha \rightarrow \infty$ 
is for the case of nearest-neighbor interaction.

For the term (\ref{Z3}) with $E(u_n,u_m)=u_m$, 
the translation invariance condition is 
\be \label{=0}
\sum_{\substack{m=-\infty \\ m \ne n}}^{+\infty} \; J(n,m)=0 
\ee
for all $n$.
If (\ref{=0}) cannot be satisfied,
we must define $E(u_n,u_m)=u_n-u_m$, and
the interaction term (\ref{Z3}) is
\be \label{Z3b}
\hat {\cal I}_{n} (u) \equiv 
\sum_{\substack{m=-\infty \\ m \ne n}}^{+\infty} \; 
J(n,m) \; [u_n-u_m] .
\ee
This interaction term is translation invariant. 
Note that the noninvariant terms lead to the divergences 
in the continuous limit (see Appendix).

In this paper, we consider the wide class of interactions (\ref{Z3b}) 
that create a possibility to present the continuous medium 
equations with fractional derivatives.
We also discuss the term (\ref{Z3}) with $E(u_n,u_m)=f(u_n)-f(u_m)$ 
as nonlinear long-range interaction. 
As the examples, we consider $f(u)=u^2$ and $f(u)=u-g u^2$ 
that give the Burgers, Korteweg-de Vries and Boussinesq equations 
and their fractional generalizations in the continuous limit.

\section{Transform operation}

In this section, we define the operation that transforms
the system of equations for $u_n(t)$ 
into continuous medium equation for $u(x,t)$. 

To derive a continuous medium equation, we 
suppose that $u_n(t)$ are Fourier coefficients
of some function $\hat{u}(k,t)$.
We define the field $\hat{u}(k,t)$ on $[-K/2, K/2]$ as 
\be \label{ukt}
\hat{u}(k,t) = \sum_{n=-\infty}^{+\infty} \; u_n(t) \; e^{-i k x_n} =
{\cal F}_{\Delta} \{u_n(t)\} ,
\ee
where  $x_n = n \Delta x$, $\Delta x=2\pi/K$ is 
distance between oscillators, and
\be \label{un} 
u_n(t) = \frac{1}{K} \int_{-K/2}^{+K/2} dk \ \hat{u}(k,t) \; e^{i k x_n}= 
{\cal F}^{-1}_{\Delta} \{ \hat{u}(k,t) \} . 
\ee
These equations are the basis for the Fourier transform, 
which is obtained by transforming 
from discrete variable to a continuous one in 
the limit $\Delta x \rightarrow 0$ ($K \rightarrow \infty$). 
The Fourier transform can be derived from (\ref{ukt}), (\ref{un}) 
in the limit as $\Delta x \rightarrow 0$.
Replace the discrete $u_n(t)$ with continuous $u(x,t)$ 
while letting $x_n=n\Delta x= 2\pi n/K \rightarrow x$.
Then change the sum to an integral, and 
Eqs. (\ref{ukt}) and (\ref{un}) become
\be \label{ukt2} 
\tilde{u}(k,t)=\int^{+\infty}_{-\infty} dx \ e^{-ikx} u(x,t) = 
{\cal F} \{ u(x,t) \}, 
\ee
\be \label{uxt}
u(x,t)=\frac{1}{2\pi} \int^{+\infty}_{-\infty} dk \ e^{ikx} \tilde{u}(k,t) =
 {\cal F}^{-1} \{ \tilde{u}(k,t) \}, 
\ee
where
\be 
\tilde{u}(k,t)= {\cal L} \hat{u}(k,t), 
\ee
and ${\cal L}$ denotes the passage 
to the limit $\Delta x \rightarrow 0$ ($K \rightarrow \infty$).
Note that $\tilde{u}(k,t)$ is a Fourier transform of the field $u(x,t)$,
and $\hat{u}(k,t)$ is a Fourier series transform of $u_n(t)$.
The function $\tilde{u}(k,t)$ can be derived from $\hat{u}(k,t)$
in the limit $\Delta x \rightarrow 0$.

The procedure of the replacement of a discrete model 
by the continuous one is defined by the transform operation. \\

{\bf Definition 1.}
{\it Transform operation $\hat T$ is a combination 
$\hat T={\cal F}^{-1} {\cal L} \ {\cal F}_{\Delta}$
of the operations: \\
1) The Fourier series transform:
\be \label{O1}
{\cal F}_{\Delta}: \quad u_n(t) \rightarrow {\cal F}_{\Delta}\{ u_n(t)\}=
\hat{u}(k,t) .
\ee
2) The passage to the limit $\Delta x \rightarrow 0$:
\be
{\cal L}: \quad \hat{u}(k,t) \rightarrow {\cal L} \{\hat{u}(k,t)\}=
\tilde{u}(k,t) .
\ee
3) The inverse Fourier transform: }
\be
{\cal F}^{-1}: \quad \tilde{u}(k,t) \rightarrow 
{\cal F}^{-1} \{ \tilde{u}(k,t)\}=u(x,t) .
\ee

The operation $\hat T={\cal F}^{-1} {\cal L} \ {\cal F}_{\Delta}$ 
is called a transform operation, since it
performs a transform of a discrete model of coupled oscillators
into the continuous medium model.  \\

{\bf Proposition 1.} 
{\it The transform operation $\hat T$ maps the function $F(u_n)$ 
into the function $F(u(x,t))$, i.e.,
\be \label{P1}
\hat T \ F(u_n(t))=F(u(x,t)) , 
\ee
where $u(x,t)=\hat T u_n(t)$,
if the function F satisfies ${\cal L} F(u_n)=F({\cal L} u_n)$. }\\

{\bf Proof}. 
The Fourier series transform leads to
\be
{\cal F}_{\Delta}: \quad F(u_n) \rightarrow {\cal F}_{\Delta} F(u_n) .
\ee
Note that ${\cal F}_{\Delta} F(u_n) \not=
F( {\cal F}_{\Delta} u_n)=F(\hat u(k,t))$.
The passage to the limit $\Delta x \rightarrow 0$ gives
\be
{\cal L}: \quad {\cal F}_{\Delta} F(u_n) \rightarrow 
{\cal L} {\cal F}_{\Delta} F(u_n) .
\ee
Then
\be
{\cal L} {\cal F}_{\Delta} \{F(u_n)\} ={\cal F} \{ {\cal L} F(u_n)\} =
{\cal F} \{ F({\cal L} u_n)\}={\cal F} \{F(u(x,t))\} ,
\ee
where we use ${\cal L} {\cal F}_{\Delta} ={\cal F} {\cal L}$. 
The inverse Fourier transform get
\be
{\cal F}^{-1}: \quad {\cal F} \{ F(u(x,t)) \} \rightarrow 
{\cal F}^{-1}\{ {\cal F}\{ F(u(x,t))\}\}=F(u(x,t)) .
\ee
As the result, we prove (\ref{P1}). \\

\section{Equations for momentum space}

Let us consider a system of infinite numbers of oscillators 
with interparticle interaction that is described by (\ref{Z3b}).
We suppose that $J(n,m)$ satisfies the condition
\be \label{Jnm}
J(n,m)=J(n-m) , \quad \sum^{\infty}_{n=1} |J(n)|^2 < \infty ,
\ee
where $J(-n)=J(n)$. \\

{\bf Proposition 2.}
{\it The Fourier series transform ${\cal F}_{\Delta}$ 
maps the equations of motion 
\be \label{C1}
\frac{\partial^2 u_n(t)}{\partial t^2}=g              
\sum^{+\infty }_{\substack{m=-\infty \\ m \not= n}}
J(n,m) [u_n-u_m]+ F(u_n) ,
\ee
where $u_n$ is the position of the $n$th oscillator,
and $F$ is an external on-site force, into the equation 
\be \label{20}
\frac{\partial^2  \hat u(k,t)}{\partial t^2}=
g [\hat{J}_{\alpha}(0)- \hat{J}_{\alpha}(k \Delta x)] \hat u(k,t) 
+{\cal F}_{\Delta} \{F(u_n)\} ,
\ee 
where   
\[ \hat{u}(k,t)={\cal F}_{\Delta}\{ u_n(t)\}, \quad
\hat{J}_{\alpha}(k \Delta x)={\cal F}_{\Delta}\{ J(n)\} , \]
and ${\cal F}_{\Delta} \{F(u_n)\}$ is an operator notation for the Fourier
series transform of $F(u_n)$.} \\

{\bf Proof.}
To derive the equation for the field $\hat u(k,t)$, we
multiply Eq. (\ref{C1}) by $\exp(-ikn \Delta x)$, 
and summing over $n$ from $-\infty$ to $+\infty$. Then
\be \label{C3a}
\sum^{+\infty}_{n=-\infty} e^{-ikn \Delta x} 
\frac{\partial^2}{\partial t^2}u_n(t)=
g  \sum^{+\infty}_{n=-\infty} \
\sum^{+\infty}_{\substack{m=-\infty \\ m \not=n}}
e^{-ikn \Delta x}  J(n,m) \ [u_n-u_m] +
\sum^{+\infty}_{n=-\infty} e^{-ikn\Delta x} F(u_n) .
\ee

From 
\be \label{C4}
\hat u(k,t)=\sum^{+\infty}_{n=-\infty} e^{-ikn \Delta x} u_n(t) ,
\ee
the left-hand side of (\ref{C3a}) gives
\be
\sum^{+\infty}_{n=-\infty} e^{-ikn \Delta x} 
\frac{\partial^2 u_n(t)}{\partial t^2}=
\frac{\partial^2 }{\partial t^2}
\sum^{+\infty}_{n=-\infty} e^{-ikn \Delta x} u_n(t)=
\frac{\partial^2 \hat u(k,t)}{\partial t^2}  .
\ee

The second term of the right-hand side of (\ref{C3a}) is
\be
\sum^{+\infty}_{n=-\infty} e^{-ikn \Delta x} F(u_n)=
{\cal F}_{\Delta} \{F(u_n)\} .
\ee

The first term of the right-hand side of (\ref{C3a}) is
defined by the function $J(n,m)$. 
Let us introduce the notation
\be \label{not}
\hat{J}_{\alpha}(k \Delta x)=
\sum^{+\infty}_{\substack{n=-\infty \\ n\not=0}} 
e^{-ikn \Delta x} J(n) .
\ee
Using $J(-n)=J(n)$, the function (\ref{not}) can be presented by
\be \label{Jcos}
\hat{J}_{\alpha}(k \Delta x)=\sum^{+\infty}_{n=1} J(n) 
\left( e^{-ikn\Delta x} +e^{ikn\Delta x} \right) = 
2\sum^{+\infty}_{n=1} J(n) \cos \left( k \Delta x \right) .
\ee
From (\ref{Jcos}) it follows that
\be
\hat{J}_{\alpha}(k \Delta x+2\pi m)= \hat{J}_{\alpha}(k \Delta x ) ,
\ee
where $m$ is an integer. 

The interaction term in (\ref{C3a}) is
\[
\sum^{+\infty}_{n=-\infty} \ \sum^{+\infty}_{\substack{m=-\infty \\ m \not=n}} 
e^{-ikn \Delta x} J(n,m) [u_n-u_m] = \]
\be \label{C6}
=\sum^{+\infty}_{n=-\infty} \  \sum^{+\infty}_{\substack{m=-\infty \\ m \not=n}}
e^{-ikn \Delta x} J(n,m) u_n - 
\sum^{+\infty}_{n=-\infty} \sum^{+\infty}_{\substack{m=-\infty \\ m \not=n}} 
e^{-ikn \Delta x} J(n,m) u_m .
\ee
Using (\ref{C4}) and (\ref{not}), 
the first term on the r.h.s. of (\ref{C6}) gives
\be \label{C7} 
\sum^{+\infty}_{n=-\infty} \ \sum^{+\infty}_{\substack{m=-\infty \\ m \not=n}}
e^{-ikn \Delta x} J(n,m) u_n =
\sum^{+\infty}_{n=-\infty} e^{-ikn \Delta x} u_n 
\sum^{+\infty}_{\substack{m^{\prime}=-\infty \\ m^{\prime} \not=0}}
J(m^{\prime})= \hat u(k,t) \hat{J}_{\alpha}(0) ,
\ee
where we use (\ref{Jnm}) and $J(m^{\prime}+n,n)=J(m^{\prime})$, and
\be \label{C8}
\hat{J}_{\alpha}(0)=\sum^{+\infty}_{\substack{n=-\infty \\ n \not=0}}
J(n)=2\sum^{\infty}_{n=1} J(n) .
\ee
For the second term on the r.h.s. of (\ref{C6}):
\[\sum^{+\infty}_{n=-\infty} \ 
\sum^{+\infty}_{\substack{m=-\infty \\ m \not=n}}
e^{-ikn \Delta x} J(n,m) u_m = 
\sum^{+\infty}_{m=-\infty} u_m 
\sum^{+\infty}_{\substack{n=-\infty \\ n \not=m}} 
e^{-ikn \Delta x} J(n,m) = \]
\be \label{C9}
=\sum^{+\infty}_{m=-\infty } u_m e^{-ikm \Delta x}
\sum^{+\infty}_{\substack{n^{\prime}=-\infty \\ n^{\prime}\not=0}} 
e^{-ikn^{\prime} \Delta x} J(n^{\prime})=
\hat u(k,t)\hat{J}_{\alpha}(k \Delta x) ,
\ee
where we use $J(m,n^{\prime}+m)=J(n^{\prime})$.

As a result, Eq. (\ref{C3a}) has the form
\be \label{C10}
\frac{\partial^2  \hat u(k,t)}{\partial t^2}=
g [\hat{J}_{\alpha}(0)- \hat{J}_{\alpha}(k \Delta x)] \hat u(k,t) 
+{\cal F}_{\Delta} \{F(u_n)\} ,
\ee 
where ${\cal F}_{\Delta} \{F(u_n)\}$ is an operator notation for the Fourier
series transform of $F(u_n)$.

\section{Alpha-interaction}

Let us consider the interaction term 
\be \label{Tuu}
\hat {\cal I}_{n} (u) \equiv 
\sum_{\substack{m=-\infty \\ m \ne n}}^{+\infty} \; 
J(n,m) \; [u_n-u_m] .
\ee
where
\be 
J(n,m)=J(n-m)=J(m-n) , 
\quad \sum^{\infty}_{n=1} |J(n)|^2 < \infty .
\ee
In Sec. 4., we prove that the Fourier series transform 
${\cal F}_{\Delta}$ of (\ref{Tuu}) gives
\be 
{\cal F}_{\Delta} \{ \hat {\cal I}_{n} (u) \}=
g [\hat{J}_{\alpha}(0)- \hat{J}_{\alpha}(k \Delta x)] \hat u(k,t) ,
\ee 
where $\hat u(k,t)={\cal F}_{\Delta}\{u_n(t)\}$, and
\be \label{Jak}
\hat{J}_{\alpha}(k)=\sum^{+\infty}_{\substack{n=-\infty \\ n\not=0}} 
e^{-ikn} J(n) = 2 \sum^{\infty}_{n=1} J(n) cos(kn) .
\ee

\noindent
{\bf Definition 2.} 
{\it The interaction term (\ref{Tuu}) in equation of motion 
(\ref{Main_Eq}) is called $\alpha$-interaction if 
the function (\ref{Jak}) satisfies the condition
\be \label{Aa}
\lim_{k \rightarrow 0} 
\frac{[\hat{J}_{\alpha}(k)- \hat{J}_{\alpha}(0)]}{|k|^{\alpha}} 
=A_{\alpha},
\ee
where $\alpha>0$ and $0<|A_{\alpha}|< \infty$.} \\

If the function $\hat{J}_{\alpha}(k)$ is given, then
$J(n)$ can be defined by
\be \label{Jnn}
J(n)=\frac{1}{\pi} \int^{\pi}_{0} \hat{J}_{\alpha}(k) cos(nk) \ dk .
\ee
The condition (\ref{Aa}) means that 
$\hat{J}_{\alpha}(k)-\hat{J}_{\alpha}(0)=O(|k|^{\alpha})$, i.e. 
\be
\hat{J}_{\alpha}(k)- \hat{J}_{\alpha}(0)=
A_{\alpha} |k|^{\alpha} +R_{\alpha}(k),
\ee
for $k\rightarrow 0$, where
\be
 \lim_{k \rightarrow 0} \ R_{\alpha}(k) / |k|^{\alpha}  =0 .
\ee

{\bf Examples}. \\

1) The first example of the $\alpha$-interaction is
\[ \hat{J}_{\alpha}(k)=A_{\alpha} |k|^{\alpha} . \]
Using (\ref{Jnn}), we obtain
\be
J(n)=A_{\alpha} \left(
\frac{(-1)^n \pi^{\alpha+1} }{\alpha+1}-
\frac{(-1)^n \pi^{1/2} }{(\alpha+1) |n|^{\alpha+1/2}}
L_1(\alpha+3/2, 1/2,\pi n) \right) ,
\ee
where $L_1(\mu,\nu,z)$ is the Lommel function \cite{Luke}. \\

2) The second example of the $\alpha$-interaction is 
\be \label{-1n2}
J(n)=\frac{(-1)^n}{n^2}  . 
\ee
Using (Ref. \cite{Prudnikov}, Sec. 5.4.2.12)
\[
\sum^{\infty}_{n=1} \frac{(-1)^n}{n^2} \cos (nk)=
\frac{1}{4}\left( k^2-\frac{\pi^2}{3} \right) , \quad |k| \le \pi ,
\]
we get
\be
\hat{J}_{\alpha}(k) =
2 \sum^{+\infty}_{n=1} \frac{(-1)^n}{n^2} cos(kn)=
\frac{1}{2} k^2- \frac{\pi^2}{6} , \quad  |k| \le \pi.
\ee
Then we have $\alpha=2$, and 
\be \label{Jk-1n2}
\hat{J}_{\alpha}(k)-\hat{J}_{\alpha}(0)=(1/2) \; k^2 .
\ee
The inverse Fourier transform of this expression
gives the coordinate derivatives of second order 
\[ {\cal F}^{-1}\{\hat{J}_{\alpha}(k)-\hat{J}_{\alpha}(0)\}=
-\frac{1}{2} \frac{\partial^2}{\partial x^2}. \]

3) For the interaction (\ref{-1n2}), we have $\alpha=2$ and 
the inverse Fourier transform of (\ref{Jk-1n2}) gives 
the second-order derivative.
At the same time, the interaction 
\[ J(n)=\frac{1}{n^2}  \] 
gives $\alpha=1$ and then the first-order coordinate derivative.
It can be proved by using 
(Ref. \cite{Prudnikov}, Sec. 5.4.2.12)
\be
\hat{J}_{\alpha}(k) =
2 \sum^{+\infty}_{n=1} \frac{cos(kn)}{n^2} =
\frac{1}{6}[3 k^2-6 \pi k+2\pi^2] ,
\quad (0\le k\le 2\pi) ,
\ee
and 
$\hat{J}_{\alpha}(k) -\hat{J}_{\alpha}(0) \approx -\pi k $
for $k \rightarrow 0$. 
Therefore, the inverse Fourier transform 
leads to the derivative of first order.\\

4) For noninteger and odd numbers $s$,
\be \label{Jnnn}
J(n)=|n|^{-(s+1)} , \quad s >0  \ee
is an $\alpha$-interaction.

For $0< s <2$ ($s \not=1$), we get 
\be 
\hat{J}_{\alpha}(k)-\hat{J}_{\alpha}(0)=
2\Gamma(-s) \cos (\pi s/ 2) \; |k|^{s} .
\ee
For $s=1$, 
\be 
\hat{J}_{\alpha}(k)-\hat{J}_{\alpha}(0)=-(\pi/2) \; k .
\ee
For noninteger $s>2$,
\be 
\hat{J}_{\alpha}(k)-\hat{J}_{\alpha}(0)=-\zeta(\alpha-1) \; k^2 ,
\ee
where $\zeta(z)$ is the Riemann zeta-function. 
The interaction ({\ref{Jnnn}}) is considered in Sec 7. \\

5) The other example is
\be
J(n)=\frac{(-1)^n}{\Gamma(1+\alpha/2+n) \Gamma(1+\alpha/2-n)} .
\ee
Using the series (Ref. \cite{Prudnikov}, Sec.5.4.8.12)
\be
\sum^{\infty}_{n=1}
\frac{(-1)^n}{\Gamma(\beta+1+n) \Gamma(\beta+1-n)} \cos(nk)=
\frac{2^{2\beta-1}}{\Gamma(2\beta+1)} 
\sin^{2\beta} \left(\frac{k}{2}\right) -\frac{1}{2\Gamma^2(\beta+1)} ,
\ee
where $\beta>-1/2$ and $0<k<2\pi$, we get
\be 
\hat{J}_{\alpha}(k)-\hat{J}_{\alpha}(0)=
\frac{2^{\alpha}}{\Gamma(\alpha+1)} 
\sin^{\alpha} \left(\frac{k}{2}\right) .
\ee
In the limit $k \rightarrow 0$, we obtain
\be \label{kka}
\hat{J}_{\alpha}(k)-\hat{J}_{\alpha}(0)
\approx \frac{1}{\Gamma(\alpha+1)} \; |k|^{\alpha} .
\ee
For noninteger $\alpha$, the inverse Fourier transform of (\ref{kka})
gives the fractional Riesz derivative \cite{SKM} of order $\alpha$. \\

6) The $\alpha$-interaction
\[ J(n)=\frac{(-1)^n}{a^2-n^2} , \]
gives
\be
\hat{J}(k)=\frac{\pi}{a\sin(\pi a)} \cos (ak) -\frac{1}{a^2} .
\ee
For $k \rightarrow 0$, we obtain
\be \label{54}
\hat{J}_{\alpha}(k)-\hat{J}_{\alpha}(0) \approx 
\frac{a \pi}{2 \sin(a \pi)} \; k^2 .
\ee
The inverse Fourier transform of (\ref{54}) leads to
the coordinate derivative of second order. \\

7) For $J(n)=1/n!$, we use
\be
\sum^{\infty}_{n=1} \frac{\cos(kn)}{n!}=e^{\cos k} \cos ( \sin k) , 
\quad |k|<\infty .
\ee
The passage to the limit $k \rightarrow 0$ gives
\be
\hat{J}_{\alpha}(k)-\hat{J}_{\alpha}(0) \approx -4e \, k .
\ee
Then $\alpha=1$, and we get the derivative of first order. \\

{\bf Proposition 3.}
{\it The transform operation $\hat T$ maps 
the discrete equations of motion
\be \label{54b}
\frac{\partial^2 u_n}{\partial t^2} = 
g  \sum_{\substack{m=-\infty \\ m \ne n}}^{+\infty} \; 
J(n,m) \; [u_n -u_m] + F (u_n) 
\ee
with noninteger $\alpha$-interaction 
into the fractional continuous medium equation: 
\be \label{CME}
\frac{\partial^2}{\partial t^2} u(x,t) -
G_{\alpha} A_{\alpha} \frac{\partial^{\alpha}}{\partial |x|^{\alpha}} u(x,t) -
F\left( u(x,t) \right) = 0  ,
\ee
where $\partial^{\alpha} / \partial |x|^{\alpha}$
is the Riesz fractional derivative, and
\be \label{G_0}
G_{\alpha}=g  |\Delta x|^{\alpha} 
\ee
is a finite parameter. } \\

{\bf Proof}. 
The Fourier series transform ${\cal F}_{\Delta}$ of (\ref{54b})
gives (\ref{20}).
We will be interested in the limit $\Delta x \rightarrow 0$. 
Then Eq.\ (\ref{20}) can be written as
\be \label{Eq-k}
\frac{\partial^2}{\partial t^2} \hat{u}(k,t) -
G_{\alpha} \; \hat{\mathcal{T}}_{\alpha, \Delta}(k) \; \hat{u}(k,t)  
-\mathcal{F}_{\Delta} \{ F \left( u_n(t) \right) \}  = 0, 
\ee
where we use finite parameter (\ref{G_0}), and 
\be 
\hat{\mathcal{T}}_{\alpha, \Delta}(k) =- A_{\alpha} |k|^{\alpha}  
-R_{\alpha} (k \Delta x) |\Delta x|^{-\alpha} .
\ee
Note that $R_{\alpha}$ satisfies the condition
\[
\lim_{\Delta x \rightarrow 0} 
\frac{R_{\alpha} (k \Delta x)}{|\Delta x|^{\alpha}} =0 .
\]
The expression for $\hat{\mathcal{T}}_{\alpha,\Delta} (k)$ can be considered
as a Fourier transform of the operator (\ref{Z3b}). 
Note that $g  \rightarrow \infty$
for the limit $\Delta x \rightarrow 0$, if $G_{\alpha}$ is a finite parameter.

In the limit $\Delta x \rightarrow 0$, Eq. (\ref{Eq-k}) gets
\be \label{Eq-k2}
\frac{\partial^2}{\partial t^2} \tilde{u}(k,t) - 
G_{\alpha} \; \hat{\mathcal{T}}_{\alpha}(k) \; \tilde{u}(k,t)  
-\mathcal{F} \{ F \left( u(x,t) \right) \}  = 0, 
\ee
where
\[ \tilde{u}(k,t)={\cal L} \hat{u}(k,t) , \quad 
\hat{\mathcal{T}}_{\alpha}(k) =
{\cal L}\hat{\mathcal{T}}_{\alpha, \Delta}(k) 
=-A_{\alpha} |k|^{\alpha} .
\]

The inverse Fourier transform of (\ref{Eq-k2}) gives
\be \label{Eq-x}
\frac{\partial^2}{\partial t^2} u(x,t) -
G_{\alpha} \; \mathcal{T}_{\alpha}(x) \; u(x,t) -
F\left( u(x,t) \right) = 0 ,
\ee
where $\mathcal{T}_{\alpha}(x)$ is an operator
\be \label{Tx0}
\mathcal{T}_{\alpha}(x) = 
\mathcal{F}^{-1} \{ \hat{\mathcal{T}}_{\alpha} (k) \} = 
A_{\alpha} \frac{\partial^{\alpha}}{\partial |x|^{\alpha}} .
\ee
Here, we have used the connection between the Riesz fractional 
derivative and its Fourier transform \cite{SKM}: 
\be
|k|^{\alpha} \longleftrightarrow - 
\frac{\partial^{\alpha}}{\partial |x|^{\alpha}}. 
\ee
The properties of the Riesz derivative 
can be found in Refs. \cite{SKM,OS,MR,Podlubny}. 
Note that the Riesz derivative could be represented as 
\be \label{N1}
\frac{\partial^{\alpha}}{\partial |x|^{\alpha}} u(x,t)=
-\frac{1}{2 \cos(\pi \alpha /2)} 
\left({\cal D}^{\alpha}_{+}u(x,t) +{\cal D}^{\alpha}_{-} u(x,t)\right),
\ee
where $\alpha \not=0,1,3,5...$, 
and ${\cal D}^{\alpha}_{\pm}$ are Riemann-Liouville 
left and right fractional derivatives defined by \cite{SKM,OS,MR,Podlubny}
\[
{\cal D}^{\alpha}_{+}u(x,t)=
\frac{1}{\Gamma(m-\alpha)} \frac{\partial^m}{\partial x^m}
\int^{x}_{-\infty} \frac{u(\xi,t) d\xi}{(x-\xi)^{\alpha-m+1}},
\]
\be \label{N2}
{\cal D}^{\alpha}_{-}u(x,t)=
\frac{(-1)^m}{\Gamma(m-\alpha)} \frac{\partial^m}{\partial x^m}
\int^{\infty}_x \frac{u(\xi,t) d\xi}{(\xi-x)^{\alpha-m+1}},
\ee
where $m-1 < \alpha < m$.

As the result, we obtain continuous medium equations (\ref{CME})
from (\ref{Eq-x}) and (\ref{Tx0}).


\section{Simple example of nearest-neighbor interaction}

In this section, we demonstrate the application
of transform operation to the well-known case:
\be
J(n,m)=\delta_{n+1, m}-2\delta_{n,m}+\delta_{n-1,m},
\ee
where $\delta_{n,m}$ is the Kronecker symbol. 
Then the interaction term (\ref{Z3}) has the form
\be
\hat {\cal I}_{n} (u) =
(u_{n+1}-u_n)-(u_n-u_{n-1}) ,
\ee
and describes the nearest-neighbor interaction.
As the result, equations of motion (\ref{C1}) have the form 
\be \label{CEM}
\frac{\partial^2 u_n}{\partial t^2} = 
g [u_{n+1}-2u_n+u_{n-1}] + F (u_n) .
\ee
The well-known result is the following. \\

{\bf Proposition 4.}
{\it The transform operation $\hat T$ maps 
the equation of motion (\ref{CEM}) 
into the continuous medium equation
\be \label{CME0}
\frac{\partial^2  u(x,t)}{\partial t^2}=
G_2 \frac{\partial^2}{\partial x^2} u(x,t) + F(u) ,
\ee
where
\be \label{GG0} G_2=g  (\Delta x)^2  \ee
is a finite parameter.} \\

{\bf Proof}.
To derive the equation for the field $\hat u(k,t)$, we
multiply Eq. (\ref{CEM}) by $\exp(-ikn \Delta x)$, 
and summing over $n$ from $-\infty$ to $+\infty$. Then
\be \label{DD1}
\sum^{+\infty}_{n=-\infty} e^{-ikn \Delta x} 
\frac{\partial^2}{\partial t^2}u_n(t)=
g  \sum^{+\infty}_{n=-\infty} \
e^{-ikn \Delta x}  [u_{n+1}-2u_n+u_{n-1}] +
\sum^{+\infty}_{n=-\infty} e^{-ikn\Delta x} F(u_n) .
\ee
The first term on the r.h.s. of (\ref{DD1}) is
\[
\sum^{+\infty}_{n=-\infty} \
e^{-ikn \Delta x}J(n,m) u_m=
\sum^{+\infty}_{n=-\infty} \
e^{-ikn \Delta x}  [u_{n+1}-2u_n+u_{n-1}] 
= \]
\[ =\sum^{+\infty}_{n=-\infty} \
e^{-ikn \Delta x}  u_{n+1} -
2 \sum^{+\infty}_{n=-\infty} \
e^{-ikn \Delta x}  u_n +
\sum^{+\infty}_{n=-\infty} \
e^{-ikn \Delta x}  u_{n-1}= \]
\[
=\sum^{+\infty}_{m^{\prime}=-\infty} \
e^{-ik(m-1) \Delta x}  u_{m} -
2 \hat{u}(k,t) +
\sum^{+\infty}_{s=-\infty} \
e^{-ik (s+1) \Delta x}  u_{s}= \]
\[ =e^{ik\Delta x} 
\sum^{+\infty}_{m^{\prime}=-\infty} \
e^{-ik m \Delta x}  u_{m} -
2 \hat{u}(k,t) +
e^{-ik \Delta x} 
\sum^{+\infty}_{s=-\infty} \
e^{-ik s \Delta x}  u_{s}=
\]
\[=e^{ik\Delta x} \hat{u}(k,t)- 2 \hat{u}(k,t) +
e^{-ik \Delta x} \hat{u}(k,t)=
[e^{ik\Delta x} +e^{-ik \Delta x}-2] \hat{u}(k,t)= \]
\be
=2[ \cos \left( k \Delta x \right)-1] \hat{u}(k,t)=
-4 \sin^2 \left( k \Delta x \right) \hat{u}(k,t) .
\ee
As the result, we obtain
\be \label{simple}
\frac{\partial^2 \hat{u}(k,t)}{\partial t^2} = 
g  \; \hat{J}_{\alpha}(k \Delta x) \; \hat{u}(k,t) +
\mathcal{F}_{\Delta} \{ F \left( u_n(t) \right) \} ,
\ee
where
\be \label{J2}
\hat{J}_{\alpha} (k \Delta x)=-4 \sin^2 \left( k \Delta x \right) .
\ee
For $\Delta x \rightarrow 0$, the asymptotics of the sine is
\[ \sin(z) =\sum^{\infty}_{m=0} \frac{(-1)^{m+1}}{(2m+1)!} z^{2m+1}
\approx z-\frac{1}{6} z^3 ,
\]
and (\ref{J2}) can be presented by
\be
\hat{J}_{\alpha} (k \Delta x)\approx 
-  \left( k \Delta x \right)^2 
+\frac{1}{12} \left( k \Delta x \right)^4  .
\ee
Using the finite parameter (\ref{GG0}), 
the transition to the limit $\Delta x \rightarrow 0$ 
in Eq. (\ref{simple}) gives 
\be \label{DD2}
\frac{\partial^2  \tilde u(k,t)}{\partial t^2}=
-G_2 k^2 \tilde u(k,t) +{\cal F}^{-1} \{F(u)\} ,
\ee
where we use $0<|G_2|<\infty$.
As the result, the inverse Fourier transform of (\ref{DD2}) leads 
to the continuous medium equation (\ref{CME0}).


\section{Integer power-law interaction}

Let us consider the power-law interaction (\ref{Z3b}) with
\be \label{J(n)0}
J(n)=|n|^{-(s+1)} 
\ee
with positive integer number $s$. \\

{\bf Proposition 5.}
{\it The power-law interaction (\ref{J(n)0})
for the odd number $s$ is $\alpha$-interaction with $\alpha=1$ for $s=1$,
and $\alpha=2$ for $s=3,5,7...$.
For even numbers $s$, (\ref{J(n)0}) is not $\alpha$-interaction. 
For odd number $s$, the transform operation $\hat T$ maps 
the equations of motion with the interaction (\ref{J(n)0}) into 
the continuous medium equation with derivatives of first order for $s=1$,
and the second order for other odd $s$. } \\

{\bf Proof}.
From (\ref{20}), we get the equation for $\hat{u}(k,t)$ in the form 
\be \label{C3bo}
\frac{\partial^2 \hat{u}(k,t)}{\partial t^2} + g  \; 
[\hat{J}_{\alpha}(k \Delta x)-\hat{J}_{\alpha}(0)] \; \hat{u}(k,t) - 
\mathcal{F}_{\Delta} \{ F \left( u_n(t) \right) \} =0,
\ee
where 
\be \label{C5o}
\hat{J}_{\alpha}(k \Delta x) = 
\sum_{\substack{n=-\infty \\ n \ne 0}}^{+\infty} 
e^{-ikn\Delta x} |n|^{-(1+s)} .
\ee
The function (\ref{C5o}) can be presented by
\be \label{C5ao}
\hat{J}_{\alpha}(k \Delta x)=\sum^{+\infty}_{n=1} \frac{1}{n^{1+s}} 
\left( e^{-ikn\Delta x} +e^{ikn\Delta x} \right) = 
2\sum^{+\infty}_{n=1} \frac{1}{n^{1+s}} \cos \left( k n \Delta x \right) .
\ee
Then we can use 
(Ref. \cite{Prudnikov} Sec. 5.4.2.12 and Sec. 5.4.2.7)
the relations
\be
\sum^{\infty}_{n=1}\frac{\cos(n k)}{n^2}=\frac{1}{12}
\left(3 k^2-6 \pi k+2 \pi^2 \right) , \quad (0\le  k \le 2\pi) ,
\ee
\be
\sum^{\infty}_{n=1}\frac{\cos(n k)}{n^{2m}}=
\frac{(-1)^{m-1} (2 \pi)^{2m}}{2(2m)!}
 B_{2m} \left( \frac{k}{2\pi} \right) ,
\quad (0\le  k \le 2\pi) ,
\ee
where $m=1,2,3,...$, and $B_{2m}(z)$ are the Bernulli polynomials \cite{BE}.
These polynomials are defined by
\be
B_{n}(k)=\sum^{n}_{s=0} C^s_n B_s k^{n-s},
\ee
where $B_s$ are the Bernoulli numbers from
\be
\frac{z}{e^z-1}=\sum^{\infty}_{s=0} B_s \frac{z^s}{s!} , 
\quad (|z|< 2\pi) .
\ee
For example,
\be
B_2(k)=k^2-k+1/6, \quad B_4(k)=k^4-2k^3+k^2-1/30 .
\ee
Note $B_{2m-1}=0$ for $m=2,3,4...$ \cite{BE}.

For $s=1$, we have
\be
\hat{J}_{\alpha}(k \Delta x)-\hat{J}_{\alpha}(0)= 
\frac{1}{2} (k \Delta x)^2 - \pi k \Delta x  
\approx - \pi k \Delta x .
\ee
For $s=2m-1$ ($m=2,3,...$), we have
\be
\hat{J}_{\alpha}(k)= \frac{(-1)^{m-1}}{(2m)!}
(2 \pi)^{2m} B_{2m} \left( \frac{k}{2\pi} \right) ,
\quad (0\le  k \le 2\pi) .
\ee
Then
\be
\hat{J}_{\alpha}(k \Delta x)-\hat{J}_{\alpha}(0)
\approx \frac{(-1)^{m-1} (2 \pi)^{2m-2}}{4(2m-2)!}  
B_{2m-2} (k \Delta x)^2 .
\ee
For example, the interaction (\ref{J(n)0}) with $s=3$ gives
\be
\hat{J}_{\alpha}(k)-\hat{J}_{\alpha}(0)= 
-\frac{1}{48} \left[ k^4-4\pi k^3+4\pi^2 k^2 \right]
\approx -\frac{\pi^2}{12} \; k^2 .
\ee

For $s=0$, we have (Ref. \cite{Prudnikov}, Sec. 5.4.2.9) the relation
\be
\sum^{\infty}_{n=1}\frac{\cos(n k)}{n}=-\ln \left[ 2 \sin (k/2) \right] .
\ee
Then, the limit $\Delta x \rightarrow 0$ gives
\be
\hat{J}_{\alpha}(k \Delta x) \approx - \ln (k \Delta x) 
\rightarrow \infty .
\ee
For even numbers $s$,
\be 
\left| \hat{J}_{\alpha}(k \Delta x)-\hat{J}_{\alpha}(0) \right| / 
\left| k \Delta x \right|^{s} \rightarrow  \infty 
\ee
since the expression has the logarithmic poles.

The transition to the limit $\Delta x \rightarrow 0$ 
in Eq. (\ref{C3bo}) with $s=1$ gives 
\be \label{DD4o}
\frac{\partial^2 \tilde{u}(k,t)}{\partial t^2} - 
G_1 \; k \; \tilde{u}(k,t) - 
\mathcal{F} \{ F \left( u(x,t) \right) \} =0,
\ee
where $G_1=\pi g  \Delta x$ is a finite parameter.
The inverse Fourier transform of (\ref{DD4o}) leads 
to the continuous medium equation with coordinate derivative 
of first order:
\be 
\frac{\partial^2}{\partial t^2} u(x,t) -
i G_{1} \; \frac{\partial}{\partial x} \; u(x,t) -
F\left( u(x,t) \right) = 0 .
\ee
This equation can be considered as 
the nonlinear Schroedinger equation.

The limit $\Delta x \rightarrow 0$ 
in Eq. (\ref{C3bo}) with $s=2m-1$ ($m=2,3,...$) gives 
\be \label{DD5o}
\frac{\partial^2 \tilde{u}(k,t)}{\partial t^2} +
G_2 \; k^2 \; \tilde{u}(k,t) - 
\mathcal{F} \{ F \left( u(x,t) \right) \} =0,
\ee
where 
\[ G_2=\frac{(-1)^{m-1} (2 \pi)^{2m-2}}{4(2m-2)!}  
B_{2m-2} \; g  (\Delta x)^2 \] 
is a finite parameter.
The inverse Fourier transform of (\ref{DD5o}) leads 
to the partial differential equation of second order:
\be \label{D10b}
\frac{\partial^2}{\partial t^2} u(x,t) -
G_{2} \; \frac{\partial^2}{\partial x^2} \; u(x,t) -
F\left( u(x,t) \right) = 0.
\ee
This equation can be considered as a nonlinear wave equation.


\section{Noninteger power-law interaction}

Let us consider the power-law interaction with
\be \label{J(n)}
J(n)=|n|^{-(s+1)} ,
\ee
where $s$ is a positive noninteger number. \\

{\bf Proposition 6.}
{\it The power-law interaction (\ref{J(n)})
with noninteger $s$ is $\alpha$-interaction
with $\alpha=s$ for $0<s<2$, and $\alpha=2$ for $s>2$.
For $0<s<2$ ($s \not=1$), 
the transform operation $\hat T$ maps the discrete equations 
with the interaction (\ref{J(n)}) into the continuous medium equation 
with fractional Riesz derivatives of order $\alpha$.
For $\alpha>2$ ($\alpha \not=3,4,5,...$), 
the continuous medium equation has the coordinate derivatives
of second order. } \\

{\bf Proof}.
From Eq. (\ref{20}), we obtain 
the equation for $\hat{u}(k,t)$ in the form 
\be \label{C3b}
\frac{\partial^2 \hat{u}(k,t)}{\partial t^2} + g  \; 
[\hat{J}_{\alpha}(k \Delta x)-\hat{J}_{\alpha}(0)] \; \hat{u}(k,t) - 
\mathcal{F}_{\Delta} \{ F \left( u_n(t) \right) \} =0,
\ee
where 
\be \label{C5}
\hat{J}_{\alpha}(k \Delta x) = 
\sum_{\substack{n=-\infty \\ n \ne 0}}^{+\infty} 
e^{-ikn\Delta x} \frac{1}{|n|^{1+\alpha}} .
\ee

For fractional positive $\alpha$,
the function (\ref{C5}) can be presented by
\be \label{C5b}
\hat{J}_{\alpha}(k \Delta x)=\sum^{+\infty}_{n=1} \frac{1}{n^{1+\alpha}} 
\left( e^{-ikn\Delta x} +e^{ikn\Delta x} \right) = 
Li_{1+\alpha}( e^{ik\Delta x} ) + Li_{1+\alpha}( e^{-ik\Delta x} ),
\ee
where $Li_{\beta}(z)$ is a polylogarithm function. 
Using the series representation of the polylogarithm \cite{Erd}:
\be \label{D1}
Li_{\beta}(e^z)=\Gamma(1-\beta) (-z)^{\beta-1}+\sum^{\infty}_{n=0}
\frac{\zeta(\beta-n)}{n!} z^n, \quad |z|< 2\pi, \; \; \beta\not=1,2,3...,
\ee
we obtain
\be \label{D2}
\hat{J}_{\alpha}(k \Delta x)= 
A_{\alpha} \; |\Delta x|^{\alpha} \; |k|^{\alpha} +
2 \sum^{\infty}_{n=0} 
\frac{\zeta(1+\alpha-2n)}{(2n)!} (\Delta x)^{2n} (-k^2)^n , 
\quad \alpha \not=0,1,2,3...,
\ee
where $\zeta(z)$ is the Riemann zeta-function, $|k \Delta x|< 2 \pi$, and
\be \label{D5}
A_{\alpha} =  2 \; \Gamma(-\alpha) \; 
\cos \left( \frac{\pi \alpha}{2} \right).
\ee
From (\ref{D2}), we have
\[ J_{\alpha}(0)=2 \zeta(1+\alpha) . \]
Then
\be  \label{96}
\hat{J}_{\alpha}(k \Delta x)-\hat{J}_{\alpha}(0)= 
A_{\alpha} \; |\Delta x|^{\alpha} \; |k|^{\alpha} +
2 \sum^{\infty}_{n=1} 
\frac{\zeta(1+\alpha-2n)}{(2n)!} (\Delta x)^{2n} (-k^2)^n ,  
\ee
where $\alpha \not=0,1,2,3...$, and $|k \Delta x|< 2\pi$. 

Substitution of (\ref{96}) into Eq. (\ref{C3b}) gives 
\[
\frac{\partial^2 \hat{u}(k,t)}{\partial t^2} + 
g  \; A_{\alpha} |\Delta x|^{\alpha} \; |k|^{\alpha} \; \hat{u}(k,t) + \]
\be \label{D4}
+2 g  \sum^{\infty}_{n=1} \frac{\zeta(\alpha+1-2n)}{(2n)!} 
(\Delta x)^{2n} (-k^2)^n \hat{u}(k,t) -
\mathcal{F}_{\Delta} \{ F \left( u_n(t) \right) \}=0 .
\ee

We will be interested in the limit $\Delta x \rightarrow 0$. 
Then Eq.\ (\ref{D4}) can be written in a simple form
\be \label{Appr}
\frac{\partial^2}{\partial t^2} \hat{u}(k,t) + 
G_{\alpha} \; \hat{\mathcal{T}}_{\alpha, \Delta}(k) \; \hat{u}(k,t)  
-\mathcal{F}_{\Delta} \{ F \left( u_n(t) \right) \}  = 0, \quad 
\alpha \not=0,1,2,...,
\ee
where we use the finite parameter
\be \label{GG}
G_{\alpha}=g  |\Delta x|^{min\{\alpha;2\}} , 
\ee
and 
\be \label{D8}
\hat{\mathcal{T}}_{\alpha, \Delta}(k) = 
\begin{cases} 
A_{\alpha} |k|^{\alpha} - |\Delta x|^{2-\alpha} \zeta (\alpha -1) k^2, 
& 0 < \alpha < 2, \quad (\alpha \not=1) ;
\cr  
|\Delta x|^{\alpha-2} A_{\alpha} |k|^{\alpha} - \zeta (\alpha -1) k^2, 
& \alpha>2 , \quad (\alpha \not=3, 4, ...).
\end{cases}
\ee
The expression for $\hat{\mathcal{T}}_{\alpha,\Delta} (k)$ 
can be considered as a Fourier transform of 
the interaction operator (\ref{Z3}). 
From (\ref{GG}), we see that $g  \rightarrow \infty$
for the limit $\Delta x \rightarrow 0$, and finite value of $G_{\alpha}$.

Note that (\ref{D8}) has a scale $k_0$: 
\be \label{D10}
k_0 = |A_{\alpha}/\zeta (\alpha-1)|^{1/(2-\alpha)} |\Delta x|^{-1} 
\ee
such that  the nontrivial expression 
$\hat{\mathcal{T}}_{\alpha, \Delta} (k) \sim |k|^{\alpha}$ 
appears only for 
$0<\alpha<2$, ($\alpha\not=1$), $k \ll k_0$. 
 
The transition to the limit $ \Delta x \rightarrow 0$ 
in Eq. (\ref{Appr}) gives
\be \label{App2}
\frac{\partial^2}{\partial t^2} \tilde{u}(k,t)+
G_{\alpha} \hat{\mathcal{T}}_{\alpha}(k) \tilde{u} (k,t)-
{\cal F}^{-1}\{ F\left( u(x,t) \right)\} =0
\quad (\alpha \not=0,1,2,...) ,
\ee
where
\be
\hat{\mathcal{T}}_{\alpha} (k) = 
\begin{cases} 
A_{\alpha} |k|^{\alpha}, & 0 < \alpha < 2, \quad \alpha \not=1; \cr 
- \zeta (\alpha -1) \; k^2, & 2< \alpha , \quad \alpha \not=3,4,....
\end{cases}
\ee

The inverse Fourier transform to (\ref{App2}) is
\be 
\frac{\partial^2}{\partial t^2} u(x,t) + 
G_{\alpha} \; \mathcal{T}_{\alpha}(x) \; u(x,t) -
F\left( u(x,t) \right) = 0 \quad 
\alpha \not=0,1,2,...,
\ee
where 
$$ \mathcal{T}_{\alpha}(x) = 
\mathcal{F}^{-1} \{ \hat{\mathcal{T}}_{\alpha} (k) \} = 
\begin{cases} 
- A_{\alpha} \; \partial^{\alpha} / \partial |x|^{\alpha} , 
& (0 < \alpha < 2, \quad \alpha \not=1); \cr 
\zeta (\alpha -1) \; \partial^2 / \partial |x|^2, 
& (\alpha >2 , \quad \alpha \not=3,4,...).
\end{cases}
$$
Here, we have used the connection between the Riesz fractional 
derivative and its Fourier transform \cite{SKM}: 
\be
|k|^{\alpha} \longleftrightarrow - 
\frac{\partial^{\alpha}}{\partial |x|^{\alpha}}, 
\quad k^2 \longleftrightarrow - \frac{\partial^2}{\partial |x|^2}.
\ee
The properties of the Riesz derivative 
can be found in \cite{SKM,OS,MR,Podlubny}.

As the result, we obtain the continuous medium equations
\be \label{D10c}
\frac{\partial^2}{\partial t^2} u(x,t) -
G_{\alpha} A_{\alpha} \frac{\partial^{\alpha}}{\partial |x|^{\alpha}} u(x,t) =
F\left( u(x,t) \right) , \quad 
0 < \alpha < 2, \quad (\alpha \not=1) ,
\ee
and 
\be \label{D10d}
\frac{\partial^2}{\partial t^2} u(x,t) +
G_{\alpha} \zeta (\alpha -1) \frac{\partial^2}{\partial |x|^2} u(x,t) =
F\left( u(x,t) \right) , \quad 
\alpha >2 , \quad (\alpha \not=3,4,...) .
\ee

Analogously, the continuous limit for the system 
\be 
\frac{\partial u_n}{\partial t} = 
g \sum_{\substack{m=-\infty \\ m \ne n}}^{+\infty} \; 
|n-m|^{-\alpha-1} \; [u_n -u_m] + F (u_n) 
\ee
gives the partial differential equations
\be \label{D10e}
\frac{\partial}{\partial t} u(x,t) -
G_{\alpha} A_{\alpha} \frac{\partial^{\alpha}}{\partial |x|^{\alpha}} u(x,t) =
F\left( u(x,t) \right) , \quad 
0 < \alpha < 2, \quad (\alpha \not=1) ,
\ee
and 
\be \label{D10f}
\frac{\partial}{\partial t} u(x,t) +
G_{\alpha} \zeta (\alpha -1) \frac{\partial^2}{\partial |x|^2} u(x,t) =
F\left( u(x,t) \right) , \quad 
\alpha >2 , \quad (\alpha \not=3,4,...) .
\ee
For $F(u)=0$, Eq. (\ref{D10e}) is
the fractional kinetic equation 
that describes the fractional superdiffusion \cite{SZ,Uch,GM2}.
If $F(u)$ is a sum of linear and cubic terms, 
then Eq. (\ref{D10e}) has the form of 
the fractional Ginzburg-Landau equation 
\cite{Zaslavsky6,TZ,TZ2,Mil,Psi}.
A remarkable property of the dynamics described by the equation with 
fractional space derivatives is that the solutions have power-like tails. 

\section{Nonlinear long-range interaction}

In this section, we consider the discrete equations
with nonlinear long-range interaction:
\be \label{nli1}
\hat {\cal I}_{n}(u)=\sum^{+\infty }_{\substack{m=-\infty \\ m \not= n}}
J_{\alpha}(n,m) [f(u_n)-f(u_m)] ,
\ee
where $f(u)$ is a nonlinear function of $u_n(t)$, and
$J_{\alpha}(n,m)$ defines the $\alpha$-interaction.
As the example of $J_{\alpha}(n,m)=J_{\alpha}(n-m)$, we can use the functions
\be \label{Js} 
J_{\alpha}(n)=\frac{(-1)^n}{\Gamma(1+\alpha/2+n) \Gamma(1+\alpha/2-n)} . 
\ee
We consider the interaction with $f(u)=u^2$ and 
$f(u)=u-g u^2$ that give the Burgers, 
Korteweg-de Vries and Boussinesq equations
in the continuous limit for $\alpha=1,2,3,4$. 
If we use the fractional $\alpha$ in Eq. (\ref{Js}), 
we can obtain the fractional generalization of these equations. \\


{\bf Proposition 7.}
{\it The Fourier series transform ${\cal F}_{\Delta}$ 
maps the equations of motion 
\be  \label{nl2}
\frac{\partial^2 u_n(t)}{\partial t^2}=g              
\sum^{+\infty }_{\substack{m=-\infty \\ m \not= n}}
J_{\alpha}(n-m) [f(u_n)-f(u_m)]+ F(u_n) ,
\ee
where $F$ is an external on-site force, into the equation 
\be \label{nle1}
\frac{\partial^2  \hat u(k,t)}{\partial t^2}=
g [\hat{J}_{\alpha}(0)- \hat{J}_{\alpha}(k \Delta x)] 
{\cal F}_{\Delta} \{f(u_n)\}+{\cal F}_{\Delta} \{F(u_n)\} ,
\ee 
where $\hat{u}(k,t)={\cal F}_{\Delta}\{ u_n(t)\}$, and 
$\hat{J}_{\alpha}(k \Delta x)={\cal F}_{\Delta}\{ J(n)\}$.

If $J_{\alpha}(n)$ defines the $\alpha$-interaction, then
the continuous limit $\Delta x \rightarrow 0$
and the inverse Fourier transform give the equation
\be \label{nle2}
\frac{\partial^2  u(x,t)}{\partial t^2}=
G_{\alpha} A_{\alpha} \frac{\partial^{\alpha} }{\partial |x|^{\alpha}}
f(u(x,t))+F(u(x,t)) ,
\ee
where $G_{\alpha}=g |\Delta x|^{\alpha}$ is a finite parameter. } \\

{\bf Proof.}
The Fourier series transform of 
the interaction term (\ref{nli1}) can be presented as 
\[ \sum^{+\infty}_{n=-\infty} \; e^{-ikn \Delta x} 
\hat {\cal I}_{n}(u)=
\sum^{+\infty}_{n=-\infty} \ 
\sum^{+\infty}_{\substack{m=-\infty \\ m \not=n}} 
e^{-ikn \Delta x} J(n,m) [f(u_n)-f(u_m)] = \]
\be \label{C6n}
=\sum^{+\infty}_{n=-\infty} \  \sum^{+\infty}_{\substack{m=-\infty \\ m \not=n}}
e^{-ikn \Delta x} J(n,m) f(u_n) - 
\sum^{+\infty}_{n=-\infty} \sum^{+\infty}_{\substack{m=-\infty \\ m \not=n}} 
e^{-ikn \Delta x} J(n,m) f(u_m) .
\ee
For the first term on the r.h.s. of (\ref{C6n}):
\be \label{C7n} 
\sum^{+\infty}_{n=-\infty} \ \sum^{+\infty}_{\substack{m=-\infty \\ m \not=n}}
e^{-ikn \Delta x} J(n,m) f(u_n) =
\sum^{+\infty}_{n=-\infty} e^{-ikn \Delta x} f(u_n) 
\sum^{+\infty}_{\substack{m^{\prime}=-\infty \\ m^{\prime} \not=0}}
J(m^{\prime})= {\cal F}_{\Delta} \{f(u_n)\} \; \hat{J}_{\alpha}(0) ,
\ee
where we use $J(m^{\prime}+n,n)=J(m^{\prime})$. 
For the second term on the r.h.s. of (\ref{C6n}):
\[\sum^{+\infty}_{n=-\infty} \ 
\sum^{+\infty}_{\substack{m=-\infty \\ m \not=n}}
e^{-ikn \Delta x} J(n,m) f(u_m) = 
\sum^{+\infty}_{m=-\infty} f(u_m) 
\sum^{+\infty}_{\substack{n=-\infty \\ n \not=m}} 
e^{-ikn \Delta x} J(n,m) = \]
\be 
=\sum^{+\infty}_{m=-\infty } f(u_m) e^{-ikm \Delta x}
\sum^{+\infty}_{\substack{n^{\prime}=-\infty \\ n^{\prime}\not=0}} 
e^{-ikn^{\prime} \Delta x} J(n^{\prime})=
{\cal F}_{\Delta} \{f(u_n)\} \; \hat{J}_{\alpha}(k \Delta x) ,
\ee
where we use $J(m,n^{\prime}+m)=J(n^{\prime})$.

As the result, we obtain Eq. (\ref{nle1}).

For the limit $\Delta x \rightarrow 0$,  
Eq.\ (\ref{nle1}) can be written as
\be 
\frac{\partial^2}{\partial t^2} \hat{u}(k,t) -
G_{\alpha} \; \hat{\mathcal{T}}_{\alpha, \Delta}(k) \; \hat{u}(k,t)  
-\mathcal{F}_{\Delta} \{ F \left( u_n(t) \right) \}  = 0, 
\ee
where we use finite parameter $G_{\alpha}=g  |\Delta x|^{\alpha}$, and 
\be 
\hat{\mathcal{T}}_{\alpha, \Delta}(k) =- A_{\alpha} |k|^{\alpha}  
-R_{\alpha} (k \Delta x) |\Delta x|^{-\alpha} .
\ee
Here, the function $R_{\alpha}$ satisfies the condition
\[ \lim_{\Delta x \rightarrow 0} 
\frac{R_{\alpha} (k \Delta x)}{|\Delta x|^{\alpha}} =0 .\]
In the limit $\Delta x \rightarrow 0$, we get
\be \label{Eq-k3}
\frac{\partial^2}{\partial t^2} \tilde{u}(k,t) - 
G_{\alpha} \; \hat{\mathcal{T}}_{\alpha}(k) \; 
\mathcal{F} \{ f \left( u(x,t) \right) \}
-\mathcal{F} \{ F \left( u(x,t) \right) \}  = 0, 
\ee
where
\[ \tilde{u}(k,t)={\cal L} \hat{u}(k,t) , \quad 
\hat{\mathcal{T}}_{\alpha}(k) =
{\cal L}\hat{\mathcal{T}}_{\alpha, \Delta}(k) 
=-A_{\alpha} |k|^{\alpha} .
\]
The inverse Fourier transform of (\ref{Eq-k3}) gives
\be 
\frac{\partial^2}{\partial t^2} u(x,t) -
G_{\alpha} \; \mathcal{T}_{\alpha}(x) \; f(u(x,t)) -
F\left( u(x,t) \right) = 0 ,
\ee
where $\mathcal{T}_{\alpha}(x)$ is an operator
\be 
\mathcal{T}_{\alpha}(x) = 
\mathcal{F}^{-1} \{ \hat{\mathcal{T}}_{\alpha} (k) \} = 
A_{\alpha} \frac{\partial^{\alpha}}{\partial |x|^{\alpha}} .
\ee

As the result, we obtain the continuous medium equation (\ref{nle2}). \\

Let us consider examples of quadratic-nonlinear 
long-range interactions. \\

1) The continuous limit of the lattice equations
\be 
\frac{\partial u_n(t)}{\partial t}=g_{1}             
\sum^{+\infty }_{\substack{m=-\infty \\ m \not= n}}
J_1(n,m) [u^2_n-u^2_m] +
g_{2} \sum^{+\infty }_{\substack{m=-\infty \\ m \not= n}}
J_2(n,m) [u_n-u_m] ,
\ee
where $J_i(n)$ ($i=1,2$) define the $\alpha_i$-interactions
with $\alpha_1=1$ and $\alpha_2=2$, 
gives the Burgers equation \cite{Burger1} that is a nonlinear partial 
differential equation of second order:
\be
\frac{\partial}{\partial t} u(x,t)+
G_1 u(x,t)\frac{\partial}{\partial x} u(x,t)-
G_2 \frac{\partial^2}{\partial x^2} u(x,t)=0 . 
\ee
It is used in fluid dynamics as 
a simplified model for turbulence, boundary layer behavior, 
shock wave formation, and mass transport.
If we consider $J_2(n,m)$ with fractional $\alpha_2=\alpha$, then
we get the fractional Burgers equation that is suggested 
in Ref. \cite{Burger2}. \\

2) The continuous limit of the system of equations
\be 
\frac{\partial u_n(t)}{\partial t}=g_{1}             
\sum^{+\infty }_{\substack{m=-\infty \\ m \not= n}}
J_1(n,m) [u^2_n-u^2_m] +
g_{3} \sum^{+\infty }_{\substack{m=-\infty \\ m \not= n}}
J_3(n,m) [u_n-u_m] ,
\ee
where $J_i(n)$ ($i=1,3$) define the $\alpha_i$-interactions
with $\alpha_1=1$ and $\alpha_3=3$, 
gives Korteweg-de Vries equation 
\be
\frac{\partial}{\partial t} u(x,t)-
G_1 u(x,t)\frac{\partial}{\partial x} u(x,t)+
G_3 \frac{\partial^3}{\partial x^3} u(x,t)=0 . 
\ee
First formulated as part of an analysis of shallow-water waves in canals, 
it has subsequently been found to be involved in a wide range of 
physics phenomena, especially those exhibiting shock waves, 
traveling waves, and solitons. 
Certain theoretical physics phenomena in the quantum mechanics 
domain are explained by means of a KdV model. 
It is used in fluid dynamics, aerodynamics, and continuum 
mechanics as a model for shock wave formation, solitons, 
turbulence, boundary layer behavior, and mass transport. 

If we use noninteger $\alpha_i$-interactions for $J_i(n)$, then 
we get the fractional generalization of the KdV equation \cite{M,M2}.\\

3) The continuous limit of 
\be 
\frac{\partial^2 u_n(t)}{\partial t^2}=
g_{2} \sum^{+\infty }_{\substack{m=-\infty \\ m \not= n}}
J_2(n,m) [f(u_n)-f(u_m)] +
g_{4} \sum^{+\infty }_{\substack{m=-\infty \\ m \not= n}}
J_4(n,m) [u_n-u_m] ,
\ee
where 
\[ f(u)=u-g  u^2 , \]
and $J_i(n)$ define the $\alpha_i$-interactions
with $\alpha_2=2$ and $\alpha_4=4$, gives
the Boussinesq equation that is a nonlinear partial differential 
equation of fourth order
\be
\frac{\partial^2}{\partial t^2} u(x,t)-
G_2 \frac{\partial^2}{\partial x^2} u(x,t)+
gG_2 \frac{\partial^2}{\partial x^2} u^2(x,t)+
G_4 \frac{\partial^4}{\partial x^4} u(x,t)=0 .
\ee
This equation was formulated as part of an analysis of 
long waves in shallow water. It was subsequently applied to problems 
in the percolation of water in porous subsurface strata. 
It also crops up in the analysis of many other physical processes.

\section{Fractional derivatives from dispersion law}

Let us consider the three-dimensional lattice that is described 
by the equations of motion
\be \label{E1} 
\frac{\partial u_{\bf n}}{\partial t} =g  
\sum_{\substack{{\bf m}=-\infty \\ {\bf m} \ne {\bf n}}}^{+\infty} \; 
J({\bf n},{\bf m} ) \; [u_{\bf n} - u_{\bf m}] + F (u_{\bf n}) ,
\ee
where ${\bf n}=(n_1,n_2,n_3)$, and 
$J({\bf n},{\bf m} )=J({\bf n}-{\bf m})=J({\bf m}-{\bf n})$.
We suppose that $u_{\bf n}(t)$ are Fourier coefficients
of the function $\hat{u}({\bf k},t)$:
\be 
\hat{u}({\bf k},t) = \sum_{{\bf n}=-\infty}^{+\infty} \; 
u_{\bf n}(t) \; e^{-i {\bf k} {\bf r}_{\bf n}} =
 {\cal F}_{\Delta} \{u_{\bf n}(t)\} ,
\ee
where ${\bf k}=(k_1,k_2,k_3)$, and
\[ {\bf r}_{\bf n}=\sum^3_{i=1} n_i {\bf a}_i . \]
Here, ${\bf a}_i$ are translational vectors of the lattice. 
The continuous medium model 
can be derived in the limit $|{\bf a}_i| \rightarrow 0$. 

To derive the equation for $\hat{u}({\bf k},t)$, 
we multiply (\ref{E1}) by $\exp(-i {\bf k} {\bf r}_{\bf n} )$,
and summing over ${\bf n}$. Then, we obtain
\be \label{E2}
\frac{\partial \hat u({\bf k},t)}{\partial t}=
g \left[ \hat{J}_{\alpha}(0)- \hat{J}_{\alpha}({\bf k} {\bf a}) \right] 
\hat u({\bf k},t) 
+{\cal F}_{\Delta} \{F(u_{\bf n})\} ,
\ee 
where ${\cal F}_{\Delta} \{F(u_{\bf n})\}$ is an operator notation 
for the Fourier series transform of $F(u_{\bf n})$, and
\be
\hat{J}_{\alpha}({\bf k} {\bf a})=
\sum_{{\bf n}=-\infty}^{+\infty} \; 
 e^{-i {\bf k} {\bf r}_{\bf n}} \; J({\bf n}) .
\ee

For the three-dimensional lattice, we define the $\alpha$-interaction 
with $\alpha=(\alpha_1,\alpha_2,\alpha_3)$,
as an interaction that satisfies the conditions:
\be \label{Aa2}
\lim_{k \rightarrow 0} 
\frac{[\hat{J}_{\alpha}({\bf k})- \hat{J}_{\alpha}(0)]}{|k_i|^{\alpha_i}} 
=A_{\alpha_i}, \quad (i=1,2,3) ,
\ee
where $0<|A_{\alpha_i}|< \infty$. 
Conditions (\ref{Aa2}) mean that 
\be
\hat{J}_{\alpha}(0)- \hat{J}_{\alpha}({\bf k})=
\sum^3_{i=1} A_{\alpha_1} |k_i|^{\alpha_i} +
\sum^3_{i=1} R_{\alpha_i}({\bf k}) ,
\ee
where
\be
\lim_{k_i \rightarrow 0} \ R_{\alpha_i}({\bf k}) / |k_i|^{\alpha_i}  =0 .
\ee
In the continuous limit ($|{\bf a}_i| \rightarrow 0$), 
the $\alpha$-interaction in the three-dimensional lattice 
gives the continuous medium equations with the derivatives
${\partial^{\alpha_1}}/{\partial x^{\alpha_1}}$, 
${\partial^{\alpha_2}}/{\partial y^{\alpha_2}}$, and 
${\partial^{\alpha_3}}/{\partial z^{\alpha_3}}$.

Let us recall the appearance of the nonlinear parabolic equation 
\cite{Leon,Light,Kad,ZS}.
Consider wave propagation in some media and 
present the wave vector $\bf k$ in the form
\begin{equation}
{\bf k} = {\bf k}_0 + {\bfkappa} = {\bf k}_0 + {\bfkappa}_{\parallel}
+ {\bfkappa}_{\perp}, 
\label{eq:36}
\end{equation}
where ${\bf k}_0$ is the unperturbed wave vector and subscripts
$(\parallel ,\perp )$ are taken respectively to the direction of ${\bf k}_0$. 
A symmetric dispersion law 
\be
\omega (k)=\omega ({\bf k})=
\hat{J}_{\alpha}({\bf k} {\bf a})-\hat{J}_{\alpha}(0)
\ee 
for $\kappa=|{\bf k}-{\bf k}_0| \ll k_0=|{\bf k}_0|$ can be written as
\be \label{PE1}
\omega (k)= \omega (|{\bf k}|) =\omega(k_0+[|{\bf k}| - k_0])
\approx 
\omega (k_0)+ \ v_g \ (|{\bf k}| - k_0 )+
{1 \over 2} v^{\prime }_g \ (|{\bf k}| - k_0 )^2 ,
\ee
where
\be \label{PE2}
v_g = \left(\frac{\partial\omega}{\partial k} \right)_{k=k_0} , \quad
v^{\prime}_g=\left(\frac{\partial^2\omega}{\partial k^2}\right)_{k=k_0} ,
\ee
and
\be \label{PE3}
|{\bf k}|=|{\bf k}_0 + {\bfkappa}|=
\sqrt{({\bf k}_0+\kappa_{\parallel})^2+\kappa^2_{\perp}} \approx
k_0+\kappa_{\parallel}+\frac{1}{2k_0}\kappa^2_{\perp}.
\ee
Substitution of (\ref{PE3}) into (\ref{PE1}) gives
\be \label{E3}
\omega (k) \approx \omega_0 + v_g\kappa_{\parallel} +
{v_g\over 2k_0} \kappa_{\perp}^2 +
\frac{v^{\prime}_g}{2} \kappa_{\parallel}^2 ,
\ee
where $\omega_0=\omega(k_0)$. 
Expressions (\ref{E2}) and (\ref{E3})  in the dual space 
("momentum representation") 
correspond to the following equation for $u=u({\bf r},t)$
in the coordinate space
\begin{equation} \label{E4}
i {\partial u \over \partial t} =\omega_0 u- 
i v_g {\partial u \over \partial x}
 - {v_g \over 2k_0 } \Delta_{\perp} u-
{v^{\prime}_g \over 2 } \Delta_{\parallel} u +F(u)
\end{equation}
with respect to the field $u=u(t,x,y,z)$, 
where $x$ is along ${\bf k}_0$, 
and we use the operator correspondence between the dual space 
and usual space-time:
\[ \omega (k)  \ \longleftrightarrow \
 i {\partial\over\partial t} , \quad
\kappa_{\parallel} \ \longleftrightarrow \ 
 -  i {\partial\over\partial x} , \] 
\be
{(\bfkappa}_{\perp})^2 \ \longleftrightarrow \ - \Delta_{\perp}=
- {\partial^2 \over\partial y^2 }-{\partial^2 \over\partial z^2 } , \quad
{(\bfkappa}_{\parallel})^2 \ \longleftrightarrow \ 
- \Delta_{\parallel}=- {\partial^2 \over\partial x^2 } .
\ee
Equation (\ref{E4}) is known as the nonlinear parabolic equation 
\cite{Leon,Light,Kad,ZS}. 
The change of variables from $(t,x,y,z)$ to $(t,x-v_gt,y,z)$ gives
\begin{equation}
-i {\partial u \over \partial t}
 = {v_g \over 2k_0} \Delta_{\perp} u +
{v^{\prime}_g \over 2} \Delta_{\parallel} u 
- \omega_0 u - F(u) ,
\end{equation}
which is also known as the nonlinear Schr\"{o}dinger equation.
 
Wave propagation in oscillatory medium with long-range interaction 
of oscillators can be easily
generalized by rewriting the dispersion law (\ref{E3}),
in the following way:
\begin{equation}
\omega (k)= \omega_0 + v_g \kappa_{\parallel} +
G_{\alpha} ({\bfkappa}_{\perp}^2 )^{\alpha /2} +
G_{\beta} ({\bfkappa}_{\parallel}^2 )^{\beta /2},
\quad (1<\alpha, \beta <2)
 \label{eq:43}
\end{equation}
with new finite constants $G_{\alpha}$, and $G_{\beta}$.

Using the connection between Riesz fractional derivative 
and its Fourier transform \cite{SKM} 
\begin{equation}
(-\Delta_{\perp} )^{\alpha /2} 
\longleftrightarrow
({\bfkappa}_{\perp}^2 )^{\alpha /2} ,
\quad
(-\Delta_{\parallel} )^{\beta /2} 
\longleftrightarrow  
({\bfkappa}_{\parallel}^2 )^{\beta /2} ,
 \label{eq:44}
\end{equation}
we obtain from (\ref{eq:43})
\begin{equation}
i {\partial u \over \partial t} =- iv_g {\partial u \over \partial x}
+ G_{\alpha} (-\Delta_{\perp} )^{\alpha /2}  u 
+ G_{\beta} (-\Delta_{\parallel} )^{\beta /2}  u 
+ \omega_0 u + F(u), 
\label{eq:45}
\end{equation}
where $u=u(t,x,y,z)$. 
By changing the variables from $(t,x,y,z)$ to $(t,\xi,y,z)$, 
$\xi=x-v_gt$, and using 
\be
(-\Delta_{\parallel} )^{\beta /2}=
\frac{\partial^{\beta} }{\partial |x|^{\beta}}=
\frac{\partial^{\beta} }{\partial |\xi|^{\beta}}, 
\ee
we obtain  from (\ref{eq:45}) 
\begin{equation}
i {\partial u \over \partial t} =
G_{\alpha} (-\Delta_{\perp} )^{\alpha /2}  u 
+ G_{\beta} (-\Delta_{\parallel} )^{\beta /2}  u 
+ \omega_0 u + F(u),
\label{eq:46}
\end{equation}
which can be called the fractional nonlinear parabolic equation.
For $G_{\beta}=0$ and $F(u)=b|u|^2 u$, we get the
fractional Ginzburg-Landau equation \cite{Zaslavsky6,TZ,TZ2,Mil,Psi}. 

We may consider one-dimensional simplifications of 
Eq. (\ref{eq:46}), i.e., 
\begin{equation}
i {\partial u \over \partial t}=
G_{\beta} \frac{\partial^{\beta} u}{\partial |\xi|^{\beta}}
+ \omega_0 u + F(u),
\label{eq:47a}
\end{equation}
where $u=u(t,\xi)$, $\xi=x-v_gt$, or 
\begin{equation}
i {\partial u \over \partial t} =
G_{\alpha} \frac{\partial^{\alpha} u}{\partial |z|^{\alpha}}
+ \omega_0 u + F(u) ,
\label{eq:47b}
\end{equation}
where $u=u(t,z)$.

Let us comment on the physical structure of (\ref{eq:46}).
The first and second terms on the right-hand side are related to 
wave propagation in oscillatory medium with long-range interaction 
of oscillators. The term with $F(u)$ on the right-hand side of
Eqs. (\ref{eq:45}), and (\ref{eq:46}) 
correspond to  wave interaction due to
the nonlinear properties of the media. 
Thus, Eq. (\ref{eq:46}) can describe
fractal processes of self-focusing and related issues.


\section{Conclusion}

One-dimensional system of long-range interacting oscillators 
serves as a model for numerous applications 
in physics, chemistry, biology, etc. 
Long-range interactions are important type of interactions 
for complex media. 
An interesting situation arises when we consider 
the wide class of $\alpha$-interactions, 
where $\alpha$ is noninteger. 
A remarkable feature of these interactions is the existence
of a transform operation that replaces the set of coupled 
individual oscillator equations by the continuous medium 
equation with the space derivative of noninteger order $\alpha$.
Such transform operation is an approximation 
that appears in the continuous limit.
This limit allows us to consider different models in a unified way 
by applying tools of fractional calculus.

Periodic space-localized oscillations, which arise in discrete 
systems, have been widely studied for short-range interactions. 
In the paper, the systems with long-range interactions were considered. 
The method to map the discrete equations of motion  
into the continuous fractional order differential equation is developed 
by the transform operation. It is known that the properties 
of a system with long-range interaction are very different 
from short-range one. The method of fractional calculus can be
a new tool for the analysis of different lattice systems.


\newpage
\section*{Appendix: Divergence of non-invatiant interaction term}

Noninvariant interaction term leads to the infinity 
in the continuous medium equation.
To demonstrate this property, we prove the following proposition.\\

{\bf Proposition 8.}
{\it The $\alpha$-interaction term
\be \label{nit}
g \sum_{\substack{m=-\infty \\ m \ne n}}^{+\infty} \; J(n,m) \; u_m ,
\ee
where $J(n,m)=|n-m|^{-(\alpha+1)}$
is not translation-invatiant. 
The transform operation $\hat T$  of the term (\ref{nit})
leads to the divergence of order $|\Delta x|^{-\alpha}$
in the continuous medium equations. } \\

Let us prove this proposition for $0< \alpha <2$ ($\alpha \not=1$),
and the following equations of motion
\be \label{ME-A}
\frac{\partial^2 u_n}{\partial t^2} + g  
\sum_{\substack{m=-\infty \\ m \ne n}}^{+\infty} \; 
J(n,m) \; u_m - F (u_n) = 0 .
\ee
Since
\[ \sum_{\substack{m=-\infty \\ m \ne n}}^{+\infty} \; J(n,m)=
\sum_{\substack{m=-\infty \\ m \ne n}}^{+\infty} \;
|n-m|^{-(\alpha+1)}  \not=0 ,\]
then the interparticle interaction term in (\ref{ME-A}) 
is noninvariant with respect to translations.
To derive the equation for $\hat{u}(k,t)$, 
we multiply Eq. (\ref{ME-A}) by $\exp(-i k n \Delta x )$,
and summing over $n$.
Then, we obtain 
\be \label{C3-A}
\frac{\partial^2 \hat{u}(k,t)}{\partial t^2} + g  \; 
\hat{J}_{\alpha}(k \Delta x) \; \hat{u}(k,t) - 
\mathcal{F}_{\Delta} \{ F \left( u_n(t) \right) \} =0,
\ee
where $\hat{J}_{\alpha}(k)$ is defined by (\ref{C5}). 
Using (\ref{D2}), we present Eq. (\ref{C3-A}) in the form
\[ \frac{\partial^2 \hat{u}(k,t)}{\partial t^2} + 
g  \; A_{\alpha} |\Delta x|^{\alpha} \; |k|^{\alpha} \; \hat{u}(k,t) + 
2 g  \zeta(\alpha+1) \hat{u}(k,t) + \]
\be \label{D4-A}
+2 g  \sum^{\infty}_{n=1} \frac{\zeta(\alpha+1-2n)}{(2n)!} 
(\Delta x)^{2n} (-k^2)^n \hat{u}(k,t) -
\mathcal{F}_{\Delta} \{ F \left( u_n(t) \right) \} =0 ,
\ee
where $\zeta$ is the Riemann zeta-function and 
$A_{\alpha}$ is defined by (\ref{D5}).
For the limit $\Delta x \rightarrow 0$ and $0<\alpha<2$ 
($\alpha\not=1$), Eq.\ (\ref{D4-A}) can be written as
\be \label{D8-A}
\frac{\partial^2}{\partial t^2} \hat{u}(k,t) + 
G_{\alpha} \; A_{\alpha} |k|^{\alpha} \; \hat{u}(k,t) + 
2 g  \zeta(\alpha+1) \hat{u}(k,t) 
-\mathcal{F}_{\Delta} \{ F \left( u_n(t) \right) \}  = 0, 
\ee
where $0 < \alpha < 2, \ \alpha \not=1$, and 
$G_{\alpha}=g  |\Delta x|^{\alpha}$ is a finite parameter. 
Note that $g  \rightarrow \infty$ for 
$\Delta x \rightarrow 0$, if $G_{\alpha}$ is a finite.
Therefore, the transition to the limit $ \Delta x \rightarrow 0$
in Eq. (\ref{D8-A}) gives the divergence term 
\be
\lim_{\Delta x \rightarrow 0}
g  \zeta(\alpha+1) \hat{u}(k,t) =
\zeta(\alpha+1) G_{\alpha} \tilde{u}(k,t) 
\lim_{\Delta x \rightarrow 0} |\Delta x|^{-\alpha} 
\rightarrow \infty .
\ee
To have the continuous model equations without divergences, 
we must consider $\left[ u_m (t)-u_n(t) \right]$ instead of 
$u_m(t)$ in the interaction terms (\ref{nit}).

\end{document}